
%
%
\magnification=1200
\hsize=5.6in
\hoffset=-.1in
\tolerance 500
\vsize=7.6in
\voffset=-.2in

\def\svec#1{\skew{-2}\vec#1}
\def\ll{\left\langle}
\def\rr{\right\rangle}
\def\onequal#1{\mathrel{\mathop{=}\limits^{#1}}}
\def\tr{\,{\rm tr}\,}
\def\Tr{\,{\rm Tr}\,}
\def\diag{\,{\rm diag}\,}

\baselineskip 12.8pt plus 1pt minus 1pt

\def\footnoterule{\kern-3pt \hrule width \hsize \kern2.6pt}
\pageno=0 \footline={\ifnum\pageno>0 \hss --\folio-- \hss \else\fi}

\centerline{\bf Test of the Skyrme Effective Field Theory}
\centerline{{\bf Using Quenched Lattice QCD}\footnote{*}
{This work is supported in part by funds provided by the U. S.
Department of Energy (D.O.E.) under contract \#DE-AC02-76ER03069.}}
\vskip 20pt
\centerline{M.--C. Chu\footnote{$^\dagger$}
{Current address:
W. K. Kellogg Radiation Laboratory,
Caltech 106-38, Pasadena CA 91125, U.S.A.}}
\vskip 6pt
\centerline{\it Center for Theoretical Physics,
Laboratory for Nuclear Science}
\centerline{\it and Department of Physics,
Massachusetts Institute of Technology}
\centerline{\it Cambridge, Massachusetts\ \ 02139\ \ \ U.S.A.}
\vskip 10pt
\centerline{Marcello Lissia}
\vskip 6pt
\centerline{\it Istituto Nazionale di Fisica Nucleare}
\centerline{\it Sezione di Cagliari}
\centerline{\it Via Ada Negri 18, I-09127 Cagliari, ITALY}
\vskip 10pt
\centerline{and}
\vskip 10pt
\centerline{J. W. Negele}
\vskip 6pt
\centerline{\it Center for Theoretical Physics,
Laboratory for Nuclear Science}
\centerline{\it and Department of Physics,
Massachusetts Institute of Technology}
\centerline{\it Cambridge, Massachusetts\ \ 02139\ \ \ U.S.A.}
\vfill
\centerline{\bf ABSTRACT}
\smallskip
\midinsert
\narrower
The Skyrme effective field theory is tested by evaluating nucleon
ground state matrix elements of the correlation functions for two
flavor density operators and two pseudoscalar density operators in the
Skyrme model and comparing them with results in quenched lattice QCD.
The possiblility of using quenched lattice QCD to study higher-order
terms in effective field theory is also discussed.
\endinsert
\vfill
\centerline{Submitted to: {\it Nuclear Physics A\/}}
\vfill
\line{CTP\#2004 \hfil hep-lat/9308012 \hfil July 1993}
\eject

\noindent{\bf I.\quad INTRODUCTION}
\medskip
\nobreak

The Skyrme model$^{(1-4)}$ provides an appealing but incomplete step
towards the goal of constructing a systematic, quantitative effective
field theory for baryons.  In the large N limit, QCD becomes
equivalent to an effective theory of mesons, and baryons emerge
naturally as solitons of this theory.  The simplest Lagrangian for the
effective theory, obtained by Skyrme by adding a non-minimal term to
the non-linear sigma model to stabilize these solitons$^1$, provides a
remarkably successful schematic model of the nucleon. However,
despite the large body of research studying specific corrections and
extensions since Adkins, Nappi and Witten$^2$ revived interest in the
Skyrme model a decade ago by proposing it as a serious model of the
nucleon, the effective field theory has never been developed and
tested in a controlled way.

Hence, the motivation for this present work is to use quenched
lattice QCD as a framework for the controlled and systematic study of
effective field theory. In contrast to studies of physical hadrons,
for which the omission of fermion loops in quenched QCD represents an
uncontrolled approximation, the fact that fermion loops are negligible
in the large N limit implies that there is a meaningful effective
meson theory for quenched QCD. Hence, lattice solutions of quenched
QCD provide an ideal laboratory to systematically test every aspect of
the approximations introduced in effective field theory.  Relative to
comparisons with experiment, the lattice QCD laboratory provides the
opportunity both to explore the approximations as a function of
fundamental parameters, such as the quark mass, and to study quantities
which are not readily amenable to experiment, such as two-body
correlation functions. Since quenched lattice QCD reproduces masses
and hadron properties which are close to experiment for the physical
quark masses, we expect the conclusions concerning effective field
theory to be relevant to physical hadrons.

To clarify the aspects of the effective field theory formulation we
would ultimately like to explore, it is useful to consider the
functional integral for full QCD. Meson fields may be introduced by
writing delta-functions setting each meson field equal to appropriate
bilinear quark fields and integrating over the meson fields. The
relevant effective action is then defined by the functional integral
over quark and gluon fields with the delta-function constraints, and
this action is to be integrated over all the meson fields,
corresponding to the calculation of all quantum loop corrections to
the stationary-phase approximation. Similarly, effective operators are
defined by the ratio of the functional integral over quark and gluon
fields of the action multiplied by the original quark and gluon
operator and delta-function constraints divided by the corresponding
integral without the operator.  In this framework, one could
systematically study the effects of truncating to various numbers of
meson fields, of truncating the terms in a derivative expansion of the
effective action, of truncating at various levels in the loop
expansion of quantum corrections, and of retaining various terms in
the expressions for effective operators. In this language, the
simplest version of the Skyrme model is obtained by retaining only the
pion field, arguing by symmetry and simplicity that only the
non-linear sigma model kinetic energy and Skyrme terms need be
retained in the action, assuming that N is large enough that quantum
corrections to the properly projected degenerate classical solutions
may be neglected, and arguing that the relevant effective operators
are uniquely specified by symmetry considerations. In principle, each
of the four classes of approximation may be tested quantitatively by
lattice calculations, affording the opportunity to explore
systematically all of the approximations underlying the effective
theory.

In this present work, we take the first step in such a program by
comparing correlation functions in the nucleon calculated on the
lattice with corresponding correlation functions calculated in the
simplest version of the Skyrme model. For each value of the quark
mass, we take the the lattice observables as defining a model system,
which we approximate by an effective Skyrme Lagrangian. As in the work
of Adkins and Nappi$^3$, we use the values of the pion, nucleon, and
delta masses to specify the three parameters in the Skyrme Lagrangian,
which then predict all other observables. Since we wish to study the
validity of the effective theory for hadron structure, we focus our
attention on two correlation functions we have calculated previously
which explore the spatial distribution of the nucleon: the
density-density correlation function and the pseudoscalar correlation
function.  Naively, since the original Skryme model with parameters
determined from hadron masses $^3$ makes a 40\% error in $F_\pi$, we
expect errors up to this order of magnitude in other observables. In
addition we expect the effective theory to improve as the pion becomes
lighter.  To the extent to which these expectations are borne out and
the effective theory provides a reasonable first approximation to the
lattice results, we believe we have established a useful framework for
quantitative investigation of each of the corrections discussed above.

The outline of this paper is as follows. In section II, we
describe the operators and correlation functions which are evaluated
on the lattice and our identification of corresponding operators in
the Skyrme model. Using the standard spin-isospin projected hedgehog
solution, analytic expressions for the correlation functions
are presented and the details of the derivations  are given
in the Appendix. The results for the density-density and pseudoscalar
correlation functions in the Skyrme model are compared with
corresponding lattice results in section III and the conclusions
are discussed in the last section.  %

\medskip
\noindent{\bf II.\quad Operators and Correlation Functions in Lattice QCD and
in the Skyrme Model}
\medskip
\nobreak

Since the QCD and the Skyrme Lagrangians are defined in
terms of different degrees of freedom, in the absence of a systematic
derivation of effective operators as discussed in the introduction,
the determination of the operators in the Skyrme model to be compared
with lattice operators might appear ambiguous.  However, at the level
of truncation considered here, we will show that the appropriate
operators may be determined from symmetry considerations.

 The Skyrme Lagrangian has been constructed such that it shares those
QCD symmetries believed to be relevant to low-energy
phenomenology: the $SU(2)_L\otimes SU(2)_R$ flavor symmetry, with its
$SU(2)_V$ subgroup that corresponds to isospin conservation, and a
topologically conserved current that corresponds to baryon number
conservation.  Since each operator of interest to us in QCD
is induced by a symmetry operation on the QCD Lagrangian, we define the
corresponding operator in the effective theory as the operator induced
by the same symmetry operation on the Skyrme Lagrangian.

\medskip
\noindent{\bf density operators}

The two flavor density operators we measure in lattice QCD are
$\hat\rho_u ({\svec r}) \equiv \colon \bar{u} ({\svec r}) \gamma^0 u
({\svec r})\colon$ and $\hat \rho_d ({\svec r}) \equiv \colon \bar{d}
({\svec r})\gamma^0 d({\svec r})\colon$. These operators may be
written as linear combinations of the time-components of the conserved
baryon current $B^\mu$ and conserved isospin current $V^{\mu,a}$ $(a =
1,2,3)$, which correspond in the standard way to the isoscalar and
isospin symmetries of QCD:
 $$\hat \rho_{u\atop d} ({\svec r}) =
{3\over 2} \hat B^0 ({\svec r}) \pm {1\over 2} \hat V^{0,3} ({\svec
r})\quad . \eqno(2.1)$$
Our normalization conventions are such that $\int d^3r\,\hat
V^{0,a}({\svec r}) = 2I_a$ ($a^{\rm th}$ component of the isospin
operator), $\int d^3r\, \hat B^0({\svec r}) = \hat B$ (baryon number
operator) and $\int d^3r\,\hat\rho_{u,d} ({\svec r}) = \hat N_{u,d}$
(flavor number operator).  Since isospin is unbroken in QCD with equal
quark masses and in the Skyrme model, both currents are exactly
conserved. Hence they do not get renormalized and the normalization in Eq.
(2.1) has an absolute meaning. Although it is possible to write
conserved currents on the lattice, the local currents we have used are
not conserved and have normalization factors which differ from unity by
the order of 10\%. As will be shown subsequently, this is not a
problem in practice since the isoscalar current strongly dominates the
isovector current.  Therefore a 10\% error in the relative normalization
is negligible and the overall density is normalized such that it
integrates to unity.

We then may write the correlation between the up and down quark density as
$$\eqalign{\ll \hat \rho_u ({\svec r}) \hat\rho_d ({\svec r}') \rr &=
{9\over 4} \ll \hat B^0 ({\svec r}) \hat B^0({\svec r}') \rr -
{1\over 4} \ll \hat V^{0,3}
({\svec r}) \hat V^{0,3} ({\svec r}') \rr \cr
&\!\!\!\onequal{\rm class.} {9\over 4} \ll\hat B^0({\svec r}) \rr \ll \hat B^0
({\svec r}') \rr - {1\over 4} \ll \hat V^{0,3} ({\svec r}) \rr \ll
\hat V^{0,3} ({\svec r}')\rr \cr
&\equiv {9\over 4} B^0 ({\svec r}) B^0 ({\svec r}') -
{1\over 4} V^{0,3}({\svec r}) V^{0,3} ({\svec r}')\ \ .\cr} \eqno(2.2)$$
While the first equality is valid in general, the second one is only valid if
we consider the classical solution and disregard the quantum fluctuations$^5$.
This is the approximation in which we solve the Skyrme model.

We therefore compare the density-density correlation function
evaluated in the nucleon ground state on the lattice
$$\ll\rho_0\rho_0\rr_L(y) \equiv \int {d^3r\,d^3r'\,\delta\left(
y -|{\svec r} - {\svec r}'|\right) \over 4\pi y^2} \ll \hat\rho_u({\svec r})
\rho_d ({\svec r}')\rr\ \ ,\eqno(2.3)$$
with the following function calculated in the Skyrme model
$$\eqalign{
\ll \rho_0\rho_0\rr_S(y) \equiv {9\over 4} &\int {d^3r\,d^3r'\,\delta \left(
y - |{\svec r} - {\svec r}'|\right)\over 4\pi y^2} B^0 ({\svec r}) B^0 ({\svec
r}') \cr
- {1\over 4} &\int {d^3r\,d^3r'\,\delta\left( y - |{\svec r} -
{\svec r}'|\right) \over
4\pi y^2} V^{0,3} ({\svec r}) V^{0,3}({\svec r}')\ \ . \cr} \eqno(2.4)$$
Here $B^\mu({\svec r})$ and $V^{\mu, 3}({\svec r})$ are the standard
currents of the Skyrme Lagrangian: $B^\mu({\svec r})$ is the
topological baryon current$^6$, and $V^{\mu,a}$ are the Noether
currents corresponding to isospin symmetry.  These currents are
evaluated for the hedgehog solution and normalized such that $\int
d^3r\,B^0({\svec r}) = 1$ and $\int d^3r\,V^{0,3} ({\svec r}) = 2I_3$.
The explicit forms of these currents, using the hedgehog solution to
the classical equation of motion, $U({\svec r}) = e^{i\hat r \cdot
{\svec\tau} F({\tilde r})}$ , are:
$$\eqalignno{
B^0({\svec r}) &= - {\left( eF_\pi\right)^3\over 2\pi^2} \ {\sin^2 F(\tilde r)
\over \tilde r^2}\ {dF(\tilde r) \over d\tilde r}&(2.5)\cr\noalign{\vskip
0.2cm}
V^{0,a} ({\svec r}) &=  F^2_\pi {\Lambda(\tilde r)\over\lambda}
\left\{ 2I_a - \left[ 3 {\svec I} \cdot \hat r_R \hat r_R -
I_a\right]\right\}\ \ ,&(2.6\hbox{a}) \cr
\Lambda(\tilde r) &\equiv {1\over 6}
\sin^2 F (\tilde r) \left\{ 1 + 4 \left[ \left(
{dF(\tilde r) \over d\tilde r} \right)^2 + {\sin^2 F(\tilde r) \over \tilde
r^2} \right]\right\}\ \ ,&(2.6\hbox{b}) \cr
\lambda&\equiv  {1\over e^3 F_\pi}
\int d^3\tilde r\,\Lambda(\tilde r)= {1\over e^3 F_\pi}\tilde \lambda \ \ .
&(2.6\hbox{c}) \cr}$$
The definition of $\lambda$ agrees with Ref.~[2] and for convenenience
we have defined the dimensionless constants $\tilde \lambda$ and
$\tilde r = e F_\pi r$.  The vector $\hat r_R$ is the unit vector in
the ${\svec r}$ direction rotated by a time-dependent angle, where
this rotation is introduced to project the hedgehog onto states of
definite spin and isospin.  Since we integrate functions that depend
only on the relative direction between $\hat r$ and $\hat r'$ over the
angles $d\Omega(\hat r)$ and $d\Omega(\hat r')$, we can always rotate
back to $\hat r,\hat r'$ and we never need to know the rotation
explicitly.  Details on how these currents and $F(\tilde r)$ are
derived are given in the Appendix.  As mentioned previously, lattice
measurements of the pion, nucleon and delta masses are used to
determine the numerical values of the three parameters that enter in
the equation for $F(\tilde r)$: the pion mass $m_\pi$, the pion decay
constant $F_\pi$ and rho-pion coupling constant $e$.

As shown in the Appendix, the final formula we obtain for nucleon
states, once angular integration has been performed is:
$$\eqalign{{1\over (eF_\pi)^3} &\ll \rho_0\rho_0\rr_{\rm
Skyrme}(\tilde y) = {9\over 4\pi^3} \ {1\over \tilde y}
\int^\infty_{{1\over 2}\tilde y} d\tilde r{\sin^2 F(\tilde r) \over
\tilde r}\ {dF(\tilde r)\over d\tilde r} \int^{\tilde r}_{|\tilde y -
\tilde r|} d\tilde r' {\sin^2 F({\tilde r}')\over \tilde r'}\
{dF(\tilde r')\over d\tilde r'} \cr &- {3\pi\over 4{\tilde\lambda}^2}
\ {1\over\tilde y} \int^{\tilde r}_{{1\over 2}\tilde y} d\tilde
r\,\tilde r\,\Lambda(\tilde r) \int^{\tilde r}_{|\tilde y - \tilde r|}
d\tilde r'\,\tilde r'\,\Lambda(\tilde r') \left[ 1+ \left( {\tilde r^2
+ \tilde r'{}^2 - \tilde y^2 \over 2\tilde r\tilde r'}\right)^2
\right] \cr} \eqno(2.7)$$
 where $\tilde y = eF_\pi y$ and we used
$(I_3)^2={1\over 4}$ for nucleon states.
\medskip
\noindent{\bf Pseudoscalar Operators}

The other operators we measured on the lattice are the pseudoscalar
density operators $\hat \rho^5_u({\svec r}) \equiv \bar{u} ({\svec r})
\gamma^5 u({\svec r})$ and $\hat \rho^5_d ({\svec r}) \equiv \bar{d}
({\svec r})\gamma^5 d({\svec r})$. For establishing the correspondence
between operators in QCD and in the Skyrme model, it is useful to note
that these operators are proportional to global chiral variations of
the QCD Lagrangian containing a mass term that explicitly breaks the
symmetry.  For notational simplicity, for a chiral transformation
$\psi\rightarrow e^{i\chi \gamma_5}\psi$ we will denote the chiral
variation $\delta\over{\delta\chi}$ by $\delta_C$. Then, disregarding
for a moment the flavor content, a chiral variation of the QCD
Lagrangian yields $\delta_{C} {\cal L}_{\rm QCD} \sim 2m_q \bar{\psi}
\gamma^5 \psi$ .  Using the fact that the ratio $m_q/m^2_\pi$ is
finite in the chiral limit (proportional to
$F_\pi^2/<\bar{\psi}\psi>$), an equivalent form for $\hat \rho^5(r)$
written purely in terms of the Lagrangian and pion mass which has a
well defined chiral limit is:
$$ \hat \rho^5(r) = {m_\pi^2\over{2m_q}}{\delta_C {\cal
L}\over{m_\pi^2}}\propto {\delta_C {\cal L}\over{m_\pi^2}}\eqno(2.8)$$
As for the flavor densities, we may write the chiral variation
as a linear combination of isoscalar and isovector terms. If we
denote the variation of the QCD Lagrangian under an isoscalar chiral
rotation by $\delta_{CS} {\cal L}_{\rm QCD} $ and the variation under
an isovector chiral rotation by $\delta_{CV} {\cal L}_{\rm QCD} $ we
may write:
$$\hat\rho^5_{u,d} = a_{u,d} m_\pi^{-2}
\delta_{CS} {\cal L}_{\rm QCD} + b_{u,d} m_\pi^{-2}
\delta_{CV}
{\cal L}_{\rm QCD} \sim a_{u,d} m_\pi^{-2}\partial_\mu \hat S_5^\mu +
b_{u,d} m_\pi^{-2}\partial_\mu
\hat V_5^{\mu,3}\ \ ,\eqno(2.9)$$
where $S_5^\mu$ and $V_5^{\mu,a}$ ($a = 1,2,3$) are the corresponding
Noether currents which are conserved in the chiral limit, and the
$a$'s and $b$'s are known constants which are finite in the chiral
limit.  We may then define the analogous operators in the Skyrme model
to within the ambiguity of the choice of the mass term, for which we
make the standard choice of ${\cal L}' = {1\over 8} m^2_\pi
F^2_\pi(\Tr U-2)$.  Since the two operators on the right-hand side of
Eq.~(2.9) are not conserved, in general one would have to calculate
the renormalized operators to determine the correct linear
combination. However, for the Skyrme Lagrangian $\delta_{CS} {\cal
L}_{\rm Skyrme} = 0$ by definition. That is, the variation is outside the
$SU(2)$ manifold where the theory is defined and there is no anomalous
contribution as for the baryon number. Hence, in the
Skyrme model both flavor operators have only the isovector
contribution
$$\left[ \hat\rho^5_u({\svec r}) \right]_{\rm Skyrme}
\sim \left[ \hat \rho^5_d ({\svec r}) \right]_{\rm Skyrme}
\sim m_\pi^{-2} \delta_{CV} {\cal L}_{\rm Skyrme}
\sim m_\pi^{-2} \partial_\mu V^{\mu,3}_{5\;\rm Skyrme} \ \ .\eqno(2.10)$$

Physically, this result corresponds to the fact that the the Skyrme
model describes the $\rho^5_u - \rho^5_d$ correlation as a pion-pion
correlation, and from PCAC we recognize $m_\pi^{-2} \partial_\mu
V_5^{\mu,3}$ as a pion interpolating field. There is no contribution
from pseudo-scalar isoscalar particles since the $\eta$ and $\eta'$
are not included in the model.  One might argue whether the model
makes a good approximation in keeping only the light pion degrees of
freedom, but once we are given this model Lagrangian, the comparison
with QCD is unambiguous. We clearly expect this approximation to
improve as we lower the pion mass.

Only the mass term ${\cal L}' = {1\over 8} m^2_\pi F^2_\pi(\Tr
U-2)$ in the Skyrme Lagrangian contributes to the chiral variation $\delta_{CV}
{\cal L}_{\rm Skyrme}$:
$$m_\pi^{-2}\delta_{CV} {\cal L}' = - {1\over 4}F^2_\pi \Tr (iU\tau_3)
= {1 \over 2}F^2_\pi\sin F(\tilde r)\hat r_R\cdot \hat z
\eqno(2.11\hbox{a})$$
where again $\hat r_R$ is the rotated vector and, once we change variables
from $\hat r$ to $\hat r_R$, we obtain
$$\rho_u ({\svec r}) \sim\rho_d ({\svec r}) \sim F^2_\pi \sin F(\tilde r)
\cos\theta \ \ .\eqno(2.11\hbox{b})$$
Note that the operator has a well defined chiral limit in the Skyrme
model.
Hence, finally we compare the correlation function evaluated in the
nucleon ground state on the lattice
$$\ll \rho_5 \rho_5\rr_L (y) \sim \int {d^3r\,d^3r'\,\delta\left( y - |{\svec
r} - {\svec r}'|\right)\over 4\pi y^2} \ll \hat \rho^5_u ({\svec r}) \hat
\rho^5_d ({\svec r}')\rr\ \ ,\eqno(2.12)$$
with the following Skyrme
model correlation function:
$$\eqalign{\ll \rho_5\rho_5\rr_S (y) &\sim \int {d^3r\,d^3r'\,\delta\left( y -
|{\svec r} - {\svec r}'|\right)\over 4\pi y^2} \sin F(\tilde r) \sin
F(\tilde r') \cos\theta
\cos\theta' \cr
&\sim {1\over y} \int^\infty_{{1\over 2}y} dr\,\sin F(\tilde r) \int^r_{|y-r|}
dr'\,\sin F(\tilde r') \left[ r^2 + r'{}^2 - y^2\right]\ \ .\cr} \eqno(2.13)$$
The overall normalization cannot be defined unambiguously as explained before,
and we can only compare the shapes of the two functions.  Details of the
angular integration are given in the Appendix.

To explore the ambiguity in the pseudoscalar current induced by the
choice of the mass term, we have considered two alternatives to the
standard result in Eq. (2.11) and compare the resulting correlation
functions in Figure 1.  Since we make the comparisons in the chiral
limit, $m_\pi = 0$, it is clear that each of the currents is
consistent with the Lagrangian and simply corresponds to a different
arbitrary choice of non-linear terms in the pion field. The standard
result from Eq. (2.11) is shown by the solid curve and labeled by
$\sin(F)$.  Variation of the mass term ${\cal L}' = {1\over 32}
m^2_\pi F^2_\pi(\Tr U^2-2)$ yields a pseudoscalar density proportional
to ${1 \over 2}\sin (2F)$ and the resulting correlation function is
shown by the dot-dashed line.  In general, any term or combination of
terms of the form ${\cal L}_k' = {1\over 8k^2} m^2_\pi
F^2_\pi\Tr(U^k-1)$ gives the expected asymptotic pion mass term and
provides a possible alternative.  The curve labeled by $F$ is obtained
by replacing $\sin F(\tilde r)$ in Eq. (2.11) by its asymptotic form at large
$\tilde r$, $F(\tilde r)$. Note that this result may be viewed as being
generated by the Lagrangian term ${\cal L}' = \sum_{k=1}^{\infty}
(-1)^k {\cal L}_k' $.  This simple comparison shows explicitly that we
can get very different results for correlation functions at short and
medium range for different choices of effective operators which are
equivalent at large distance. This clearly motivates future efforts to
derive or calculate on the lattice higher-order terms in the expansion
of effective operators.

In a previous comparison$^7$ of the lattice results and the bag model,
it was natural to equate the quark field operators of the two
theories. In the present context, one may ask whether the results
would have been different if we had used operators induced by
symmetries.  It is readily seen that whereas we would have obtained the
same flavor density operators, the pseudoscalar density operators
would have been somewhat different.  The main point is that in
contrast to the QCD and Skyrme Lagrangians, the bag Lagrangian in the
basic form used in ref.~[7] is not invariant under chiral rotation,
even in the limit of zero quark mass.  In the chiral limit, the chiral
variation receives non-vanishing contributions from the bag surface.
This contribution is equal in form to the QCD term, but it is
multiplied by a delta function at the surface.  Hence, in addition to
the double volume integral of the relevant quark wave functions, we
would  also have had a double surface integral with a weight that would
have become dominant in the chiral limit. The shape of this latter
contribution is significantly different, even though it retains some
of the correct global features such as having zero volume
integral$^7$.

\goodbreak
\bigskip
\noindent{\bf III.\quad RESULTS}
\medskip
\nobreak

We use quenched lattice QCD calculations at three different values of the bare
quark mass to define three test systems which may then be approximated by the
Skyrme model.  The lattice calculations are carried out on a $16^3 \times 24$
lattice at $\beta=5.7$ using bag model sources and are described in detail in
Refs.~[7-10]. We have chosen to define the energy scale of the lattice
calculations using the string tension, which gives an inverse lattice spacing
$ a^{-1}=1$GeV and $a=0.2$fm.  Although defining the scale using the
extrapolated proton mass yields a value of $a$ 15\% smaller, the quantitative
comparison with the Skyrme model is unaffected since when the Skyrme
parameters are determined from the lattice masses, the correlation functions
scale in $a$.

The equation for the soliton, Eq.~(A.5), depends only on the
dimensionless parameter $\beta = {m_\pi\over{eF_\pi}}$ and its
solution is a function of the dimensionless variable $\tilde r =
eF_\pi r$. Therefore, the correlation functions we calculate depend
only on two parameters, $m_\pi$ and the scale factor $(eF_\pi)^{-1}$.
However, since we link the choice of $eF_\pi$ to the proton and delta
masses through Eq.~(A.3), we actually determine $e$ and $F_\pi$
separately.  The four sets of Skyrme parameters we use are shown in
Table I, where the masses in bold face have been used as input and the
other parameters are predicted by the model. Since the nucleon and
delta energies depend nonlinearly on $m_\pi\over{eF_\pi}$, the
parameters are determined numerically. A rough idea of the sensitivity
of the nucleon or delta mass to $eF_\pi$ is given by the fact that at
$m_\pi = 0$, ${\delta M \over M} \sim 2{\delta(eF_\pi)\over{eF_\pi}}$

\medskip
\noindent {\bf density-density correlation function}
\smallskip

As a prelude to considering the Skyrme model density-density
correlation function's dependence on the quark mass and how it
compares with the lattice results, it is useful to note the extent to
which it is dominated by the isoscalar baryon number density
contribution.  The isovector and isoscalar contributions, as well as
the weighted average ${9\over8} <\rho\rho>_S - {1\over8} <\rho\rho>_V$ are
shown in Figure 2 for the range of pion masses we consider.  In all
cases, over the spatial range shown in the figure relevant to this
work, one observes that the isoscalar and isovector contributions are
sufficiently similar that $1\over8$ of their difference is very small
compared to the isoscalar contribution and the full correlation
function is very well approximated by the isoscalar correlation
function.  This result has two important implications. We may regard
the physics as being dominated by the topological baryon number
density and, as claimed earlier, we are justified in neglecting
differences in the renormalization factors for the isovector and
isoscalar density operators on the lattice. The only place the
isovector and isoscalar correlation functions differ significantly is
in the extreme surface, where the isoscalar correlation decays
asymptotically as
$e^{-3m_\pi r}$,
whereas the isovector correlation function decays as
$e^{-2m_\pi r}$. This behavior reflects the leading
contributions of two or three pions respectively in each channel and
gives rise to the fact that at very large distances, the correlation
function changes sign.  Similarly, we also expect the correlation to
be negative at sufficiently large distance on the lattice.

The change in the shape and spatial extent of the correlation function
with quark mass is shown in Figure 3.  For convenience, throughout
this section we will refer to pion mass dependence interchangeably with
quark mass dependence. It is implicit that when we consider a change
in the pion mass, we also consider the correlated changes in $F_\pi$
and $e$ induced when a change in the bare quark mass produces this
change in the pion mass.

A striking feature of Fig.~3 is the fact that the mass dependence is highly
non-linear.  Indeed, the change in the correlation function caused by reducing
$m_\pi$ by 173 MeV from 515 to 342 MeV is quite small compared to the change
caused by reducing $m_\pi$ by another 205 MeV to 137 MeV.  A related
indication of how far these masses are from the chiral limit is given by the
values shown in Table I for the parameter $\beta = {m_\pi \over e \, F_\pi}$
governing the hedgehog solution.

The radial extent of the correlation function
depends both on the scale factor $(eF_\pi)^{-1}$, which decreases
as $m_{\pi}$ decreases, and on the exponential tail, which grows as
$m_{\pi}$ decreases. These two effects go in opposite directions for
the cases we consider. It turns out that the scale factor dominates
and the net result is shrinking of both the isoscalar and isovector
contributions with decreasing $m_\pi$.  This shrinkage with decreasing
$m_\pi$ is shown both by the increase of the normalized correlation
function at zero separation and by the decrease in isoscalar and
isovector r.m.s. radii in Table I.  (Note that for values of $m_\pi$
lighter than the physical mass, the isovector radius begins to
increase and ultimately diverges in the chiral limit because of chiral
logs.) The trend is opposite in the lattice results, where the
normalized correlation function at zero separation clearly decreases
with $m_\pi$.  The second moments of the lattice results have
sufficiently large errors because of the $r^2$ weighting of the poorly
determined tails that their trend with $m_\pi$ is not statistically
determined.  Also note that these second moments of correlation
functions are not rms radii of density distributions because of the
presence of polarization contributions$^{11}$.

Lattice measurements of the density-density correlation function for each of
the three values of the bare quark mass are compared with the corresponding
Skyrme model correlation functions in Figure 4.  Because statistical errors in
the correlation functions at large radius induce significant errors in the
overall normalization, we have normalized all correlation functions to unity
at the origin to facilitate comparison.  One observes that as the quark mass
and thus the pion mass decreases, the agreement improves substantially.  Since
we are using sufficiently large quark masses that we should not expect the
Skyrme model to be an accurate approximation, it is encouraging that the
agreement improves so conspicuously as the quark mass decreases.  It will
therefore be of interest to carry out more detailed comparisons for lighter
quark masses, to verify that this agreement is maintained as the correlation
function undergoes the significant changes shown in Fig.~3.

\medskip
\noindent {\bf pseudoscalar correlation function}
\smallskip

In contrast to the density-density correlation function, where both
isoscalar and isovector terms contribute, the Skyrme model has only an
isovector contribution to the pseudoscalar correlation function.  This
correlation function decays exponentially in the pion mass at large
$r$ and, on physical grounds, it should be the dominant contribution
in the chiral limit at large distance.  In contrast, the lattice
calculation has contributions from both the isoscalar and isovector
terms. At the large quark masses we consider in this work, the
isovector $\pi$ and the isoscalar $\eta$ have comparable masses and
may be expected to yield comparable contributions with an overall
coefficient that favors the isoscalar one. Hence, we expect the best
agreement between the Skyrme and lattice results to occur near the
chiral limit, where the isovector contribution becomes dominant.

The dependence of the spatial distribution of the pseudoscalar
correlation function on the quark mass is shown in Figure 5. Although
in principle one has the same competition between the shrinking scale
parameter and growing pion tail as in the density-density correlation
function, we observe that the pseudoscalar correlation function has
extremely mild mass dependence in the range 0-2 fm.  The only
qualitative feature which is not shown in this plot is the fact that
in the Skyrme model, the volume integral of this correlation function
is zero, as it is in any independent particle approximation$^7$.
Thus, the function has a node and an exterior negative region of equal
volume.  When the correlation function is plotted at larger distances,
one observes that as the pion mass becomes lower, the tail extends to
larger distances and the node moves outward.

The comparison of the Skyrme model pseudoscalar correlation functions
with the lattice results is shown in Figure 6. On the lattice we
observe that the overall size is very different, as expected because
of the large isoscalar contribution at the quark masses we have used.
However, it is significant that as the pion mass decreases, the
lattice result becomes much closer to the Skyrme solution.
The outward movement of the node is particularly evident in Figure 6.
The stability of the Skyrme correlation function when varying the mass
compared to the large change of the lattice result suggests that a
significant part of change in the lattice result arises from the
diminishing contribution of the isoscalar term.  Again it would be
very interesting to follow the behavior to lower quark masses.  As a
technical point, we note that because of the node at large distance
and the slow spatial decay at small pion mass, the pseudoscalar
correlation function is more sensitive to finite size effects on
the lattice, and is more difficult to extrapolate to the physical pion
mass than the density-density correlation function.

\goodbreak
\bigskip
\noindent{\bf IV.\quad SUMMARY AND CONCLUSIONS}
\medskip
\nobreak

In this work, we have derived the analytic results necessary to compare the
density-density and pseudoscalar correlation functions calculated in lattice
QCD with the Skyrme model and taken the first step in using quenched lattice
QCD as a laboratory to systematically explore the Skyrme effective field
theory.  The principal numerical results are presented in Figures 4 and 6.  We
observe in Figure 4 that already for a quark mass of 40 MeV, corresponding to
a pion mass of 340 MeV, the Skyrme approximation to the density-density
correlation function appears to be in good agreement with the lattice result.
In contrast, because of substantial isoscalar contributions in quenched QCD
for a pion mass of 340 MeV, which are excluded by construction from the Skyrme
model, the Skyrme pseudoscalar correlation function is not quantitatively
correct, although the trend with decreasing pion mass is encouraging.

These exploratory results suggest several promising directions for
future research.  At the most pedestrian level, it will clearly be
worthwhile to extend the comparison to lighter quark masses.  More
interesting and potentially extremely useful, is the opportunity to
use lattice techniques to calculate numerically the effective action
and effective operators and thus  evaluate corrections to the Skyrme
approximation.  As described in the introduction, this provides a
consistent framework to systematically explore the quantitative
effects of truncation in meson fields, of truncation of the effective
action, of truncation of the expansion of effective operators, and of
truncation of the quantum loop corrections.  Thus, rather than using
the lattice to calculate hadron observables directly, one can instead
use it to calculate from first principles the relevant parameters of
an effective theory which could then be applied much more generally
than one could hope to apply the full apparatus of lattice QCD.

\bigskip
\noindent{\bf Acknowledgments}

It is a pleasure to acknowledge illuminating discussions with Suzhou
Huang and Janos Polonyi. The authors also gratefully acknowledge the
hospitality of the Institute for Nuclear Theory, where this work and
manuscript were completed, and the supercomputer resources for the
lattice calculations provided by the National Energy Research
Supercomputer Center.

\vfill
\eject
\centerline{\bf REFERENCES}
\medskip
\item{1.}T. H. R. Skyrme, {\it Proc. Roy. Soc.\/} {\bf A260} (1961)
127.
\medskip
\item{2.}G. S. Adkins, C. R. Nappi and E. Witten, {\it Nucl.
Phys.\/} {\bf B228} (1983) 552.
\medskip
\item{3.}G. S. Adkins and C. R. Nappi, {\it Nucl.
Phys.\/} {\bf B233} (1984) 109.
\medskip
\item{4.}I. Zahed and G. E. Brown, {\it Physics Reports\/} {\bf 142}
(1986) 1.
\medskip
\item{5.}S. Huang, {\it Nucl. Phys.\/} {\bf B324} (1989) 34.
\medskip
\item{6.}J. Goldstone and F. Wilczek, {\it Phys. Rev. Lett.\/} {\bf
47} (1981) 986.
\medskip
\item{7.}M. Lissia, M.-C. Chu, J. W. Negele and J. M. Grandy,{\it
Nucl. Phys.\/} {\bf A555} (1993) 272.
\medskip
\item{8.}M.-C. Chu, M. Lissia and J. W. Negele, {\it Nucl. Phys.\/}
{\bf B360} (1991) 31.
\medskip
\item{9.}M.-C. Chu, J. M. Grandy, M. Lissia and J. W. Negele,
{\it Nucl. Phys.\/} {\bf B (Proc. Suppl.) 26} (1992) 412.
\medskip
\item{10.}J. Grandy, Ph. D. Thesis (1992) M.I.T., Cambridge, MA.
\medskip
\item{11.} M. Burkardt, J. M. Grandy, and J. W. Negele, MIT preprint
CTP\#2109 (1993).
\medskip
\item{12.}P. Bacilieri {\it et. al., Nucl. Phys.\/} {\bf B317} (1989) 509.

\vfill
\eject
\centerline{\bf APPENDIX}
\bigskip

In this Appendix we review the main features of the Skyrme model, we
give an explicit derivation of the time component of the vector
current without averaging over angular variables (which is also useful for
other
applications) and we perform the angular integration when the two
currents are kept at fixed relative distance.  As in
the rest of the paper we have tried to keep notation consistent with
Refs.~[2,3], from which Eqs.~(A.1) through (A.7) are taken.

The Lagrangian of the model is (Ref.~[3])
$${\cal L} = {1\over 16} F^2_\pi \tr\left[\partial_\mu U \partial^\mu
U^\dagger\right] + {1\over 32e^2} \tr\left[\left[ \left( \partial_\mu U\right)
U^\dagger, \left( \partial_\mu U\right) U^\dagger\right]^2 \right] + {1\over
8} m^2_\pi F^2_\pi (\tr U-2)\ \ .\eqno(\hbox{A.1})$$
If we substitute the spinning hedgehog {\it ansatz\/}
$$\eqalign{ U&= A(t) U_0 ({\svec r}) A^\dagger(t)\ \ ,\qquad A(t) \in SU(2) \cr
U_0({\svec r}) &= e^{iF(\tilde r) {\svec\tau}\cdot\hat r} \cr}
\eqno(\hbox{A.2a})$$
into (A.1), we obtain after quantization the following expression for the
energy
(Ref.~[2]):
$$E = {F_\pi\over e} \left[ M + {e^4 \over 8\tilde\lambda} 2J (2J+2)\right]
\eqno(\hbox{A.3})$$
where $J$ is the spin (equal to isospin in this model) of the particle
$$M = \int d^3 \tilde r \left\{ {1\over 8}\left[ \left( {dF\over d\tilde
r}\right)^2 + 2{\sin^2 F\over \tilde r^2} \right] + {1\over 2} \left[ 2 \left(
{dF\over d\tilde r}\right)^2 + {\sin^2 F\over \tilde r^2} \right] {\sin^2
F\over \tilde r^2} + {1\over 4} \beta^2 (1-\cos F)\right\}\ \
,\eqno(\hbox{A.4})$$
$\tilde\lambda$ was defined in Eq.~(2.6c), $\tilde r = eF_\pi r$ and $\beta =
{m_\pi\over eF_\pi}$.  $F(\tilde r)$ is chosen so that it minimizes $M$ with
the
boundary conditions $F(0) = \pi$ and $F(\infty) = 0$ to ensure that the baryon
number (integral of Eq.~(2.5)) is equal to 1; the resulting equation for
$F(\tilde r)$ is
$$\eqalign{\left( {1\over 4} \tilde r^2 + 2\sin^2 F\right) {d^2 F\over d\tilde
r^2} &+ 2 \sin F\cos F \left( {dF\over d\tilde r}\right)^2 + {1\over 2} \tilde
r
{dF\over d\tilde r} \cr
&-2\sin F\cos F\left( {1\over 4} + {\sin^2 F\over \tilde r^2} \right)
-{1\over 4} \beta^2 \tilde r^2 \sin F = 0\ \ .\cr} \eqno(\hbox{A.5})$$

The general form of the time component of the Noether current associated with
the $V-A$ transformation $\delta U = i\tau_a U$ of the Lagrangian (A.1) is
(Ref.~[2]):
$$\eqalign{J^{0,a}_{V-A} &= {1\over 8} i F_\pi^2 \tr\left\{
\left(\partial^0U\right)U^\dagger \tau_a\right\} + {i\over 8e^2} \tr\left\{
\left[ \left( \partial_\nu U\right)U^\dagger, \tau_a\right] \left[ \left(
\partial^0 U\right) U^\dagger,\left(\partial^\nu U\right) U^\dagger\right]
\right\} \cr
&= {1\over 8} i F_\pi^2 \tr\left\{ \left( \partial^0 U\right) U^\dagger
\tau_a\right\} - {i\over 8e^2} \Sigma_i \tr\left\{ \left[ \left( \partial_i
U\right) U^\dagger,\tau_a\right] \left[ \left( \partial^0 U \right)
U^\dagger, \left( \partial_i U\right) U^\dagger\right] \right\}\ \ .\cr}
\eqno(\hbox{A.6})$$
By using $\Tr \left\{ [A,B][C,D] \right\} = 2\Tr \{AD\} \Tr \{BC\} - 2 \Tr \{
AC\} \Tr \{ BD\}$, valid for $A,B,C,D$ belonging to the $SU(2)$ Lie algebra
and the fact that $\partial_\mu U^\dagger = \partial_\mu U^{-1} = - U^{-1}
\left(\partial_\mu U\right) U^{-1} = - U^\dagger\left( \partial_\mu U\right)
U^\dagger$, we may rewrite (A.6) as
$$\eqalign{J^{0,a}_{V-A} &= {iF^2_\pi\over 8} \tr\left\{ \left(\partial^0
U\right) U^\dagger \tau_a\right\} - {i\over 4e^2} \tr\left\{\left( \partial^0
U\right) U^\dagger \tau_a \right\} \Sigma_i \tr\left\{\left[ \left( \partial_i
U\right) U^\dagger\right]^2 \right\} \cr
&\quad + {i\over 4e^2} \Sigma_i \tr\left\{ \left( \partial_i U \right)
U^\dagger
\tau_a\right\} \tr \left\{ \left( \partial_i U \right)U^\dagger\left(
\partial_0 U\right) U^\dagger\right\} \cr
&= {iF^2_\pi\over 8} \tr\left\{ \left( \partial_0U\right) U^\dagger
\tau_a\right\} + {i\over 4e^2} \tr\left\{ \left( \partial^0 U\right) U^\dagger
\tau_a\right\} \Sigma_i \tr\left\{ \left( \partial_i U\right) \left( \partial_i
U^\dagger\right)\right\} \cr
&\quad - {i\over 4e^2} \Sigma_i \tr\left\{ \left( \partial_i U\right)
U^\dagger\tau_a \right\} \tr \left\{ \left( \partial_i U\right) \left(
\partial_0 U^\dagger\right) \right\}\ \ .\cr} \eqno(\hbox{A.7})$$
Before calculating the four traces we need, let us derive a few useful
formulae.  Equation (A.2a) can be written as
$$\eqalign{U_0({\svec r}) &= \cos F + i {\svec\tau}\cdot \hat r \sin F \cr
U({\svec r}, t) &= A(t) U_0 A^\dagger(t) = \cos F + i {\svec\tau} \cdot \hat
r_R \sin F \cr} \eqno(\hbox{A.2b}) $$
where ${\svec r}_R$ is the vector ${\svec r}$ rotated according to the law
${\svec r}_R \cdot {\svec\tau} = A{\svec r} \cdot {\svec\tau} A^\dagger$.
Using the fact that $A$ is unitary, we obtain
$${dU\over dt} = \dot U = \dot A U_0 A^\dagger + AU_0 \dot A^\dagger = \dot A
A^\dagger A U_0 A^\dagger - A U_0 A^\dagger\dot A A^\dagger = \left[ \dot A
A^\dagger, U\right]\ \ .\eqno(\hbox{A.8})$$
The spatial derivatives are
$$\eqalign{
{\partial U_0 \over \partial x_i} &= \partial_i U_0 =  \hat r_i {dF\over dr}
\left( - \sin F + i \cos F\hat r \cdot {\svec\tau}\right) + i {\sin F\over
r}\left( \tau_i - \hat r_i \hat r \cdot {\svec \tau}\right) \cr
{\partial U\over\partial x^i_R} &= \partial^R_i U =  \hat r_{Ri} {dF\over dr}
\left( - \sin F + i \cos F \hat r_R \cdot {\svec\tau}\right) + i {\sin F\over
r} \left( \tau_i - \hat r_{Ri} \hat r_R \cdot {\svec\tau}\right) \ \ .\cr}
\eqno(\hbox{A.9})$$
We also need to show that
$$\Sigma_i\left[ \partial_i (\ldots) \right] \left[ \partial_i(\ldots)\right] =
\Sigma_i \left[\partial^R_i (\ldots) \partial^R_i (\ldots)\right]
\eqno(\hbox{A.10a})$$
or equivalently that
$$J_{jk} \equiv \Sigma_i {\partial r_{Rj}\over \partial r_i}\ {\partial
r_{Rk}\over \partial r_i} = \delta_{jk}\ \ .\eqno(\hbox{A.10b})$$
This is because it is more convenient to take derivatives of $U$ with respect
to ${\svec r}_R$ (see Eq.~(A.9)).  Since
$${\partial r_{Rj}\over \partial r_i} = {\partial\over\partial r_i}\ {1\over
2} \tr\left\{ {\svec r}_R\cdot {\svec\tau}\tau_j\right\} =
{\partial\over\partial r_i}\ {1\over 2} \tr\left\{ {\svec r}\cdot {\svec\tau}
A^\dagger\tau_j A\right\} = {1\over 2} \tr\left( \tau_i A^\dagger \tau_j
A\right)\ \ .$$
we may write
$$\eqalign{J_{jk} &= {1\over 4}\Sigma_i \tr\left\{ \tau_i \left( A^\dagger
\tau_j A\right)\right\} \tr
\left\{ \tau_i \left( A^\dagger\tau_k A\right)\right\} \cr
&= \Sigma_i {1\over 2} \left\{ \left( A^\dagger\tau_j A\right) \left(
A^\dagger\tau_k A\right)\right\} = {1\over 2} \tr\left\{ \tau_j \tau_k\right\}
= \delta_{jk}\ \ ,\cr} \eqno(\hbox{A.10c}) $$
where we used the fact that
$${1\over 2} \tr\left\{ M_1\, M_2\right\} = {1\over 2} \tr \left\{ M_1\right\}
{1\over 2} \tr\left\{ M_2\right\} + \Sigma_i {1\over 2} \tr\left\{ \tau_i
M_1\right\} {1\over 2} \tr\left\{ \tau_i M_2\right\}\ \ ,\qquad M_1, M_2 \in
SU(2)\ \ .\eqno(\hbox{A.11})$$
This is the component form of the scalar product in $SU(2)$.  Now
we are ready to calculate the traces in Eq.~(A.7).  The first one gives
$$\eqalign{\tr \left\{\left( \partial_0 U\right) U^\dagger \tau_a\right\} &=
\tr\left\{ \left[ \dot AA^\dagger,U\right] U^\dagger \tau_a \right\} \cr
&= \tr\left\{ \left[ \dot AA^\dagger,i\sin F\hat r_R\cdot{\svec\tau}\right]
\left( \cos F - i \sin F\hat r_R \cdot {\svec\tau}\right) \tau_a\right\} \cr
&= i \sin F\cos F\tr\left\{\left[
\dot AA^\dagger,\hat r_R \cdot {\svec \tau}\right]
\tau_a \right\} +\sin^2 F\tr\left\{ \left[ \dot A A^\dagger,\hat
r_R\cdot{\svec\tau}\right] \hat r_R \cdot {\svec\tau}\tau_a\right\} \cr
&=i \sin (2F) \tr\left\{ \dot A A^\dagger \hat r_R
\cdot {\svec\tau} \tau_a\right\}
+ 2\sin^2 F\tr\left\{ \dot A A^\dagger\tau_a\right\} \cr
&\quad - \sin^2 F \tr\left\{ \dot A A^\dagger\hat r_R \cdot
{\svec\tau}\right\} \tr\left\{ \hat{r}_R\cdot {\svec\tau} \tau_a\right\}\ \
,\cr} \eqno(\hbox{A.12})$$
where we used $\tr\left\{ [A,B]CD\right\} = \tr \{AD\} \tr \{BC\} - \tr\{
AC\} \tr\{ BD\}$ and $\tr\{ABC\} = - \tr\{BAC\}$ for $A,B,C,D,$ belonging to
the $SU(2)$ Lie algebra to go from the second last to the last line.  The
second trace is
$$\eqalign{\Sigma_i \tr \left\{ \left( \partial_i U\right)\left( \partial_i
U^\dagger\right)\right\} &= \Sigma_i
\tr\left\{ \left( \partial_i U_0\right) \left(
\partial_i U^\dagger_0\right)\right\} \cr
&= \Sigma_i \tr \left\{ \left[ \hat r_i {dF\over dr}\left( - \sin F + i\cos F
\hat r \cdot {\svec\tau}\right) + i {\sin F\over r} \left( \tau_i - \hat r_i
\hat r \cdot{\svec\tau}\right)\right]\right. \cr
&\quad \times \left.\left[ \hat r_i {dF\over dr}
\left( - \sin F - i \cos F\hat r \cdot
{\svec \tau}\right) - i {\sin F\over r} \left( \tau_i -\hat r_i  \hat r
\cdot {\svec \tau}\right)\right] \right\} \cr
&= \tr\left\{ \left( {dF\over dr}\right)^2 + {\sin^2 F\over r^2} \left(
{\svec\tau}^2 - 2 + 1 \right)\right\} = 2\left[\left( {dF\over dr}\right)^2 +
2 {\sin^2 F\over r^2}\right]\ \ .\cr} \eqno(\hbox{A.13})$$
The third trace is, using the fact that in the last term of Eq.~(A.7) we shall
take advantage of property (A.10a) to change $\partial_i$ into
$\partial^R_i$,
$$\eqalign{\tr\left\{ \left(\partial^R_i U\right)U^\dagger \tau_a\right\} &=
\tr\biggl\{ \left[ \hat r_{Ri} {dF\over dr}\left( - \sin F + i\cos F \hat r_R
\cdot {\svec\tau}\right) + i {\sin F\over r} \left( \tau_i - \hat r_{Ri} \hat
r_R\cdot {\svec\tau}\right)\right] \cr
&\quad \times \left[ \cos F - i \sin F\hat r_R \cdot {\svec\tau}\right]
\tau_a\biggr\} \cr
&= \hat r_{Ri} \tr\left\{ \ldots\right\} + i {\sin F\over r} \Tr \left\{\left[
\cos F - i \sin F\hat r_R\cdot {\svec \tau}\right] \tau_a \tau_i\right\}\ \
.\cr} \eqno(\hbox{A.14})$$
We do not write out the term proportional to $\hat r_{Ri}$, because it will
not contribute when multiplied times the fourth trace, which is
$$\eqalign{\tr\left\{ \left( \partial_0 U\right)\left( \partial^R_i
U^\dagger\right)\right\} &= \tr\left\{ \left[ \dot A A^{-1}, U \right]
\partial^R_i U^{-1}\right\} = \tr \left\{ \dot AA^{-1} \left[ U,\partial^R_i
U^{-1}\right]\right\} \cr
&= \tr\left\{ \dot A A^{-1} \left[ i \sin F\hat r_R\cdot {\svec\tau}, - i
{\sin F\over r} \tau_i\right]\right\} \cr
 &= {\sin^2 F\over r} \tr\left\{ \dot A A^{-1} \left[\hat r_R \cdot
{\svec\tau} ,\tau_i\right]\right\} = 2{\sin^2 F\over r} \tr\left\{ \dot A
A^{-1} \hat r_R \cdot {\svec \tau} \tau_i\right\} \ \ .\cr}
\eqno(\hbox{A.15})$$
Now we multiply Eq.~(A.14) times Eq.~(A.15) and sum over $i$ to get the last
term of Eq.~(A.7). Note that the first term in Eq.~(A.14) when multiplied
times Eq.~(A.15) gives a result
proportional to $\tr \left\{ \dot A A^{-1} \left(\hat r_R\cdot {\svec
\tau}\right)^2\right\} = \tr\left\{ \dot A A^{-1}\right\} = 0$. To
combine the trace in the second term with the one in Eq.~(A.15) we may use
Eq.~(A.11):
$$\eqalign{\Sigma_i &\tr\left\{ \left( \partial_i U\right) U^\dagger
\tau_a\right\} \tr\left\{ \left( \partial_i U\right)\left(
\partial_0U^\dagger\right)\right\} \cr
&= \Sigma_i \tr\left\{ \left( \partial^R_i U\right) U^\dagger\tau_a\right\}
\tr\left\{ \left( \partial^R_i U\right)\left( \partial_0 U^\dagger\right)
\right\} \cr
&= i 8 {\sin^3F\over r^2} {1\over 2} \tr \left\{ \dot AA^\dagger\hat r_R
\cdot {\svec\tau}\tau_i \right\} {1\over 2} \tr \left\{ \left[ \cos F - i \sin
F \hat r_R\cdot {\svec\tau}\right] \tau_a\tau_i\right\} \cr
&= i 8 {\sin^3F\over r^2} \left( {1\over 2} \tr \left\{
\dot AA^\dagger \hat r_R\cdot
{\svec\tau}\left[ \cos F - i \sin F\hat r_R \cdot {\svec\tau}\right]
\tau_a\right\}\right.\cr
&\quad\left. - {1\over 2} \tr\left\{ \dot AA^\dagger\hat r_R \cdot
{\svec\tau}\right\} {1\over 2} \tr\left\{ - i\sin F\hat r_R \cdot {\svec\tau}
\tau_a\right\}\right) \cr
&= 2{\sin^2F\over r^2} \left[ i \sin 2F\tr\left\{ \dot AA^\dagger \hat r_R
\cdot {\svec\tau} \tau_a\right\} + 2\sin^2 F \tr\left\{ \dot AA^\dagger
\tau_a\right\} \right.\cr
&\quad\left.- \sin^2 F \tr\left\{ \dot AA^\dagger  \hat r_R \cdot
{\svec\tau}\right\} \tr \left\{ \hat r_R \cdot {\svec\tau} \tau_a\right\}
\right] \ \ .\cr} \eqno(\hbox{A.16})$$
The expression in brackets is equal to $\tr\left\{\left(
\partial_0 U\right)U^\dagger \tau_a\right\}$ (Eq.~(A.12)), which appears in
both the other two terms in Eq.~(A.7). We then  insert Eq.~(A.12), Eq.~(A.13)
and Eq.~(A.14) into Eq.~(A.7) and obtain
$$\eqalign{J^{0,a}_{V-A} &= {iF^2_\pi\over 4} \left[ \sin^2 F \tr\left\{ \dot
AA^{\dagger} \tau_a\right\} + {i\over 2}
\sin 2F \tr\left\{ \dot AA^\dagger\hat r_R \cdot
{\svec\tau} \tau_a\right\} - \hat r_{Ra} \sin^2 F \tr\left\{ \dot
AA^\dagger\hat
r_R\cdot {\svec\tau}\right\}\right] \cr
&\quad \times \left[ 1 + {4\over (eF_\pi)^2}\left(\left( {dF\over dr}\right)^2
+
{\sin^2 F\over r^2}\right)\right]\ \ .\cr} \eqno(\hbox{A.17})$$
As in Ref.~[2], one can convince oneself that $J^{0,a}_{V+A}$ is
obtained from
Eq.~(A.7) by exchanging $U\leftrightarrow U^{-1}$, {\it i.e.\/} $\hat r_R
\leftrightarrow - \hat r_R$. Then  $J_V$ will contain terms even in $\hat r_R$
and
$J_A$ terms odd in $\hat r_R$, as expected from parity considerations:
$$\eqalign{ J^{0,a}_{V} &= J^{0,a}_{V+A} + J^{0,a}_{V-A}
={iF^2_\pi\over 2} \sin^2 F \left[ 1 + {4\over
(eF_\pi)^2} \left( \left( {dF\over dr}\right)^2 + {\sin^2 F\over r^2}\right)
\right] \cr
&\quad
\times \left[ \tr\left\{ \dot AA^{-1} \tau_a\right\} - \hat r_{Ra} \tr\left\{
\dot AA^\dagger \hat r_R\cdot {\svec\tau}\right\} \right] \cr
&= 3
i F^2_\pi \Lambda(\tilde r) \left[ \tr \left\{ \dot AA^{-1} \tau_a\right\}
- \hat r_{Ra} \tr\left\{\dot AA^\dagger\hat r_R\cdot {\svec\tau}\right\}
\right] \cr
&= F^2_\pi {\Lambda(\tilde r)\over{\lambda}} \left\{ 2I_a -  \left[
3{\svec I}\cdot \hat r_R \hat r_{Ra} - I_a\right]\right\}\ \ ,
\cr}\eqno(\hbox{A.18})$$
where we used the definitions of $\Lambda$ and $\lambda$, Eqs.~(2.6b) and
(2.6c).  We also used the fact that $\dot AA^{-1} = - {i\over 2\lambda} {\svec
I}\cdot
{\svec\tau}$ with ${\svec I}$ the isospin operator, as may be
derived from $A=a_0 + i {\svec a}\cdot {\svec\tau}$, ${\svec\pi} =
4\lambda\dot
{\svec a}$ and ${\svec I} = - {1\over 2} \left[ a_0{\svec\pi} - {\svec
a}\pi_0 - {\svec a} \times {\svec\pi}\right]$ which can be found in Ref.~[2].
For completeness, the time component of the axial current is
$$J^{0,a}_A = J^{0,a}_{V+A} - J^{0,a}_{V-A}
= {F^2_\pi\over 4\lambda} \sin 2F \left\{ 1 + {4\over (eF_\pi)^2}
\left[ \left( {dF\over dr}\right)^2 + {\sin^2F\over r^2} \right]\right\}
\left( {\svec I} \times \hat r_R \right)_a\ \ ,\eqno(\hbox{A.19})$$
in agreement  with Ref.~[3].

We calculate only the correlations between currents in proton (or neutron),
$\ll N|VV|N\rr$, and not off-diagonal matrix elements, $\ll
N|VV|N'\rr$. Hence,
we only need:
$$\eqalign{\diag &\left[ J^{0,3}_V ({\svec r}) J^{0,3}_V ({\svec r}')\right]
 = F^4_\pi
{9\Lambda(\tilde r)\Lambda(\tilde r') \over \lambda^2} \diag\left[ \left( I_3
- \bar{I} \cdot \hat r \hat r_3 \right)\left( I_3 - {\svec I}\cdot \hat r'
\hat r'_3 \right)\right] \cr
&= F^4_\pi {9\Lambda(\tilde r)\Lambda (\tilde r') \over \lambda^2}\left\{
(I_3)^2 - \hat r_i \hat r_3 \diag (I_i I_3) - \hat r'_i r'_3 \diag (I_i I_3) +
\hat r_i \hat r_3 \hat r'_j \hat r'_3 \diag (I_i I_j\right\} \cr
&= F^4_\pi {9\Lambda(\tilde r)\Lambda(\tilde r')\over \lambda^2} (I_3)^2
\left\{ 1 - \cos^2 \theta - \cos^2\theta' + \cos\gamma\cos\theta\cos\theta'
\right\}\ \ ,\cr} \eqno(\hbox{A.20})$$
where $\gamma$ is the angle between ${\svec r}$ and ${\svec r}'$.

The correlation we want is then (note that we changed variables of integration
from ${\svec r}, {\svec r}'$ to ${\svec r}_R, {\svec r}'_R$ and that
$(I_3)^2={1\over 4}$ both in the proton and neutron):
$$\eqalign{
V(y) &= \int {d{\svec r}\,d{\svec r}'\,\delta(y-|{\svec r}- {\svec
r}'|)\over 4\pi y^2} \diag \left[ J^{0,3}_V ({\svec r}) J^{0,3}_V ({\svec r}')
\right] \cr
&={9\over 4}\ {1\over 4\pi y^2}\ {1\over (e^3 F_\pi \lambda)^2} \
\int^\infty_0 d\tilde r\,\tilde r^2 \int^\infty_0 d\tilde r'\,\tilde r'^2
\Lambda(\tilde r) \Lambda(\tilde r')
\int d\phi \int d\phi'
\cr
&\quad \times \int d(\cos\theta) \int d(\cos\theta')
\delta(y-|{\svec r} - {\svec r}'|) \left[ 1 - \cos^2\theta - \cos^2\theta' +
\cos\gamma \cos\theta\cos\theta'\right] \cr} \eqno(\hbox{A.21})$$
Now we use
$$\delta\left( y - |{\svec r} - {\svec r}'|\right) = 2y \,\delta\left( y^2 -
( {\svec r} - {\svec r}')^2\right) = {\tilde y eF_\pi\over\tilde r\tilde r'}
\delta\left( \cos\gamma - {\tilde r^2 + \tilde r'{}^2 - \tilde y^2\over
2\tilde r \tilde r'}\right)\ \ ,\eqno(\hbox{A.22})$$
where $\tilde y = eF_\pi$;
we change variables from $\phi,\theta,\phi',\theta'$ to
$\phi,\theta,\Phi,\gamma$, where $\Phi$ and $\gamma$
are the angles of ${\svec r}'$ with
respect to ${\svec r}$ (the transformation is just a rotation and its
Jacobian is one); and we use the following identities
$$\eqalignno{\cos\theta' &= \cos\theta\cos\gamma \pm \sin\theta
\sin\gamma\cos(\phi\pm\Phi)\ \ ,&(\hbox{A.23a}) \cr
\int d\phi\cos\theta' &= 2\pi\cos\theta\cos\gamma\ \ ,&(\hbox{A.23b}) \cr
\int d\phi \cos^2 \theta' &= 2\pi\left[\cos^2\theta\cos^2\gamma + {1\over
2}\sin^2\theta\sin^2\gamma\right]\ \ ,&(\hbox{A.23c})\cr}$$

Note that we need only (A.23b) and (A.23c) that can be derived from (A.23a)
without knowing the signs which depend on conventions.  Then,
$$\eqalign{
V(y) &= {9\over 4}\ {1\over (e^3 F_\pi\lambda)^2}\
{(e F_\pi)^3\over 4\pi\tilde y}\
\int^\infty_0 d\tilde r\,\tilde r\,\Lambda(\tilde r) \int^\infty_0 d\tilde
r'\,\tilde r'\,\Lambda(\tilde r') (2\pi)^2 \int d(\cos\theta) \int
d(\cos\gamma)\cr
&\quad \times \delta \left( \cos\gamma  - {\tilde r^2 + \tilde r'{}^2 - \tilde
y^2 \over 2\tilde r\tilde r'}\right) \cr
&\quad \times \left[ 1 - \cos^2\theta - \cos^2\theta \cos^2\gamma
- {1\over 2}( 1 -
\cos^2\theta)(1-\cos^2\gamma) + \cos^2\theta\cos^2\gamma\right] \cr
&= {9\pi\over 4}  {1\over(e^3F_\pi\lambda)^2}\ {(eF_\pi)^3\over \tilde
y}
\int^\infty_0 d\tilde r\,\tilde r\,\Lambda(\tilde r) \int^\infty_0 d\tilde
r'\,\tilde r'\,\Lambda(\tilde r') \cr
&\quad\times \int^1_{-1} dz\,\delta
\left( z - {\tilde r^2
+ \tilde r'{}^2 - \tilde y^2\over 2\tilde r \tilde r'} \right) {1\over 2}
(1+z^2) \int^1_{-1} dx(1-x^2) \cr
&= {3\pi\over 2} \ {(eF_\pi)^3\over \tilde y}\ {1\over(e^3 F_\pi\lambda)^2}
\int^\infty_0 d\tilde r\,\tilde r\,\Lambda(\tilde r) \int^\infty_0 d\tilde
r'\,\tilde r'\,\Lambda(\tilde r') \cr
&\quad \times \left[ 1 + \left( {\tilde r^2 + \tilde r'{}^2
- \tilde y^2\over 2\tilde r\tilde r'}\right)^2\right]\theta\left( 2\tilde r
\tilde r' - |\tilde r^2+\tilde r'{}^2 - {\tilde y}^2|\right) \cr
&= 3\pi\ {(eF_\pi)^3\over \tilde y}\ {1\over (e^3 F_\pi\lambda)^2}
\int^\infty_{{1\over 2}\tilde y} d\tilde r\,\tilde r\,\Lambda(\tilde r)
\int^{\tilde r}_{|\tilde r - \tilde y|} d\tilde r'\, \tilde r'\,\Lambda(\tilde
r') \left[ 1 + \left( {\tilde r^2 + \tilde r'{}^2 - \tilde y^2 \over 2rr'
}\right)^2 \right]\ \ ,\cr} \eqno(\hbox{A.24})$$
where in the last line we use the fact that the integrand is symmetric in
$\tilde r$ and $\tilde r'$ to impose $\tilde r \ge \tilde r'$ and multiply
the result times two; with this condition $2\tilde r \tilde r' \ge |\tilde
r^2 + \tilde r'{}^2 - \tilde y^2|\Leftrightarrow \tilde r' \ge |\tilde r -
\tilde y|$ which in turn implies $\tilde r > {1\over 2} \tilde y$.  The other
two angular integrals we need can be derived in the same way:
$$\int {d{\svec r}\,d{\svec r}' \,\delta(y - | {\svec r} - {\svec r}'|) \over
4\pi y^2} F(r) F(r') = {4\pi\over y} \int^\infty_{{1\over 2} y} dr\,r\,F(r)
\int^r_{|y-r|} dr'\,r' F(r') \eqno(\hbox{A.25})$$
and
$$\eqalign{\int &{d{\svec r}\,d{\svec r}'\,\delta(y - |{\svec r} - {\svec r}'|)
\over 4\pi y^2} F(r) F(r') \cos\theta\cos\theta '\cr
&= {1\over 3} \int {d{\svec
r}\,d{\svec r}' \,\delta(y - |{\svec r} - {\svec r}'|) \over 4\pi y^2} F(r)
F(r') \cos\gamma \cr
&= {4\pi\over 3y} \int^\infty_{{1\over 2} y} dr\,r\,F(r) \int^r_{|y-r|}
dr'\,r'\,F(r') \left( {r^2 + r'{}^2 - y^2\over 2rr'}\right)\ \
.\cr}\eqno(\hbox{A.26})$$
Note that the limit for vanishing $y$ of each of the last three equations is
finite$^7$.
\par
\vfill
\eject

\baselineskip 16pt plus 1pt minus 1pt 

\centerline{\bf Figure Captions}
\medskip
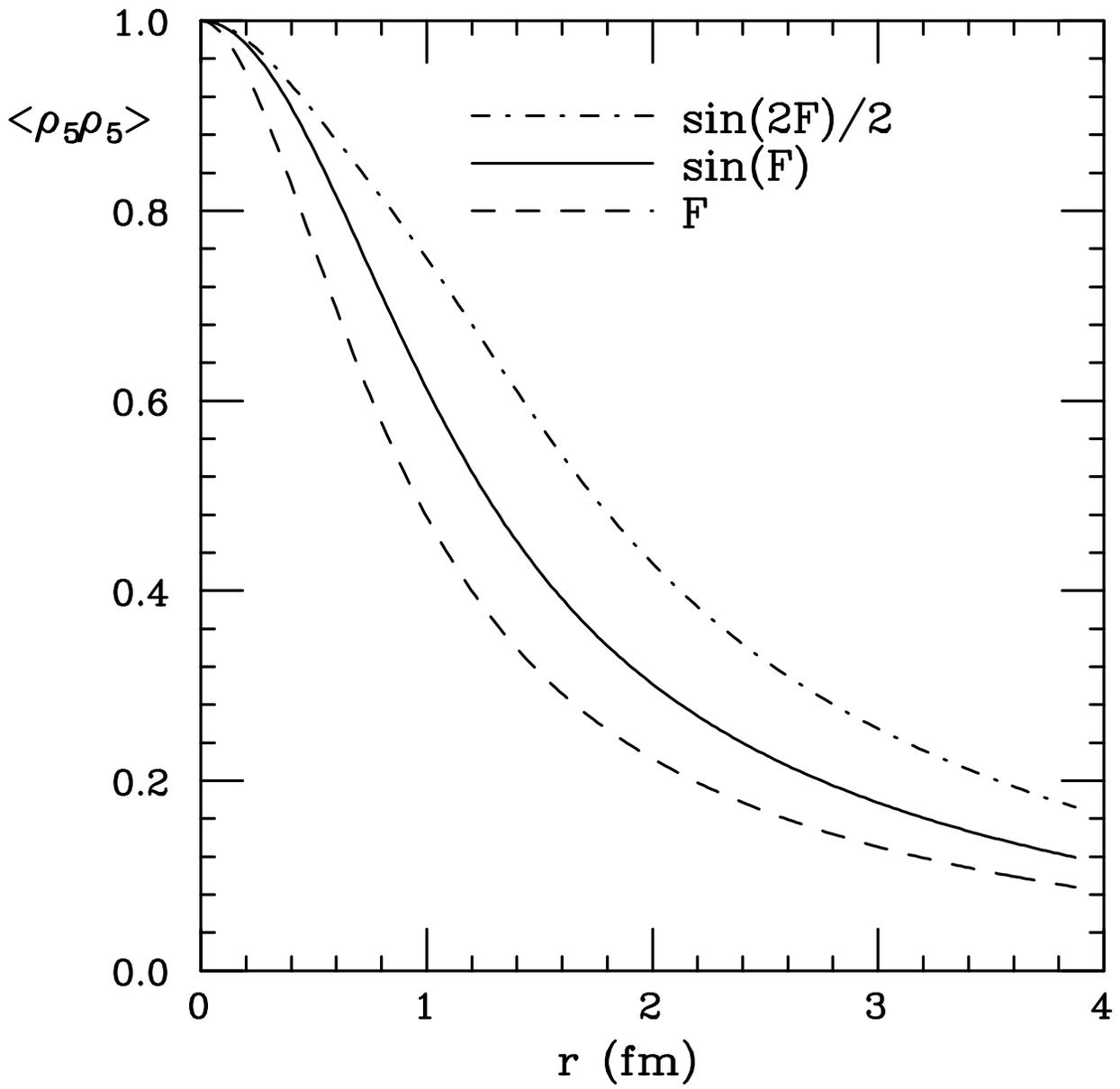
\item{Fig. 1} Pseudoscalar correlation function in the Skyrme model for
three alternative definitions of the pseudoscalar current.  The solid
curve denotes the standard current, Eq.~(2.11b), induced by the usual
choice of the mass term. The dashed and dot-dashed curves show two
other alternatives discussed in the text.
\medskip
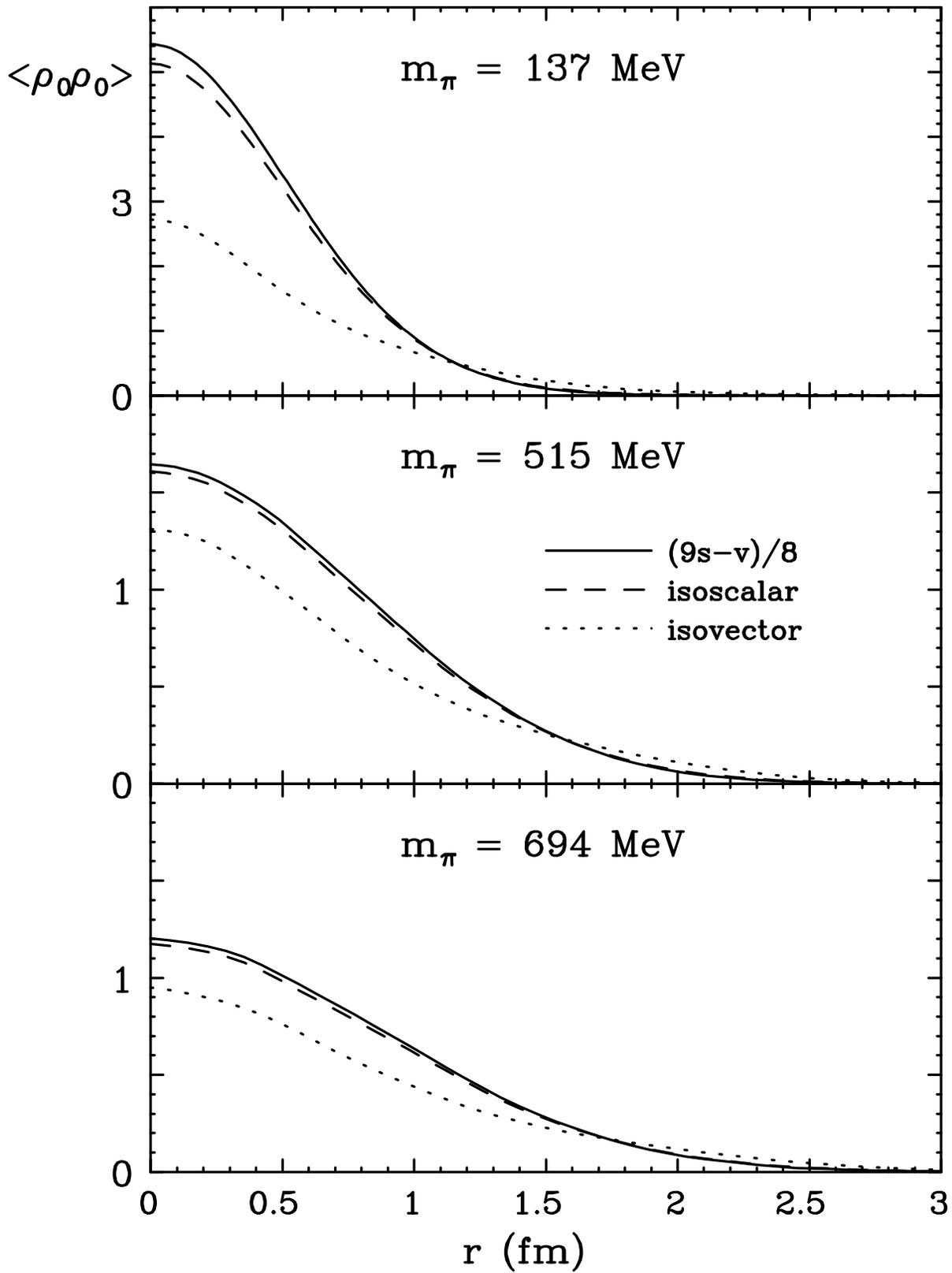
\item{Fig. 2} Relative contribution of isoscalar and isovector
contributions to the density-density correlation function in the
Skyrme model.  Dashed and dotted lines show the isoscalar and isovector
contributions respectively at three representative values of the quark
mass, and the weighted combination, Eq.~(2.2), is shown by the solid
curve.
\medskip
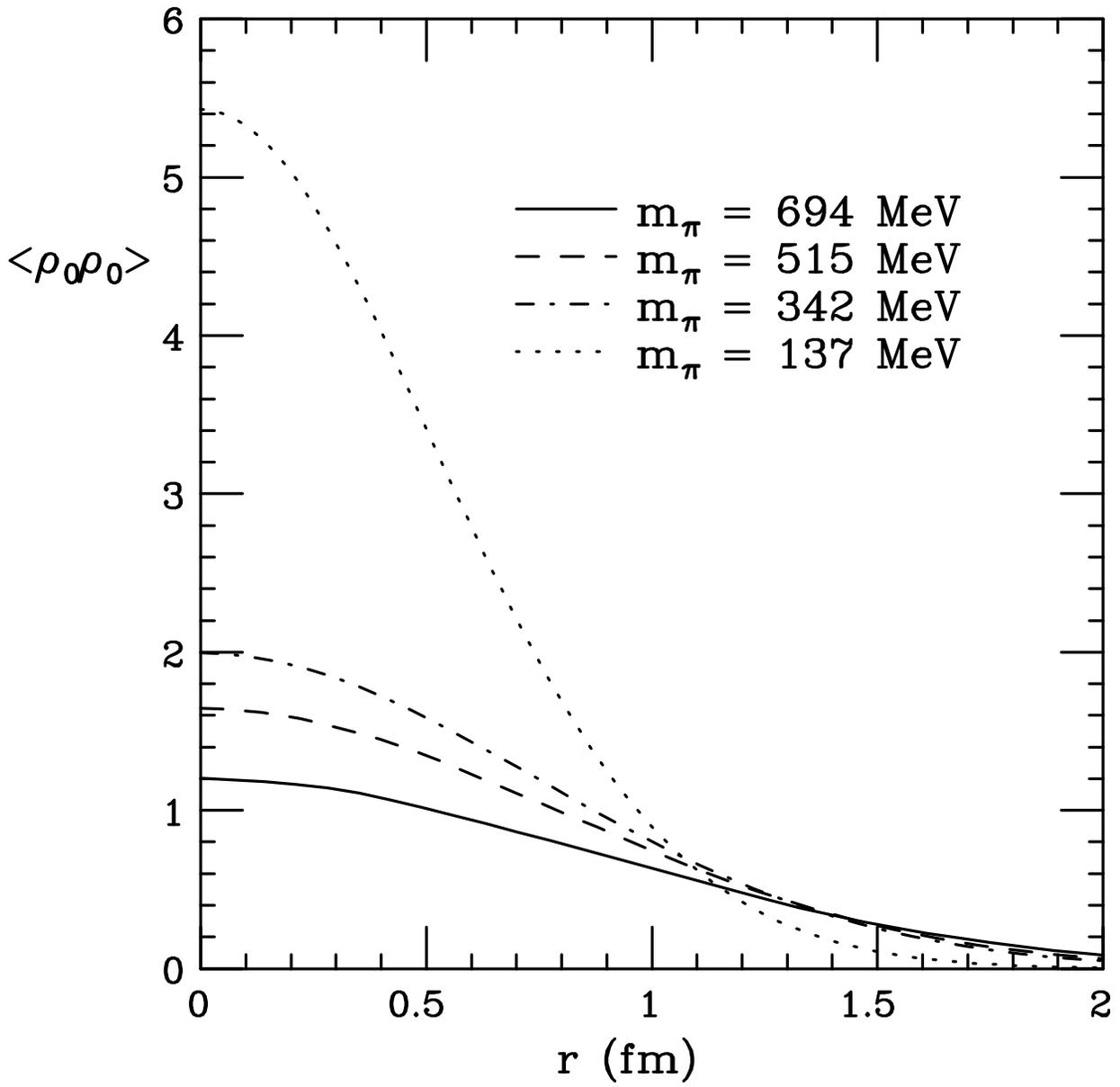
\item{Fig. 3} Mass dependence of the density-density correlation
function in the Skyrme model.  The spatial distribution of the
correlation function is shown for each of the four sets of parameters
in Table I.
\medskip
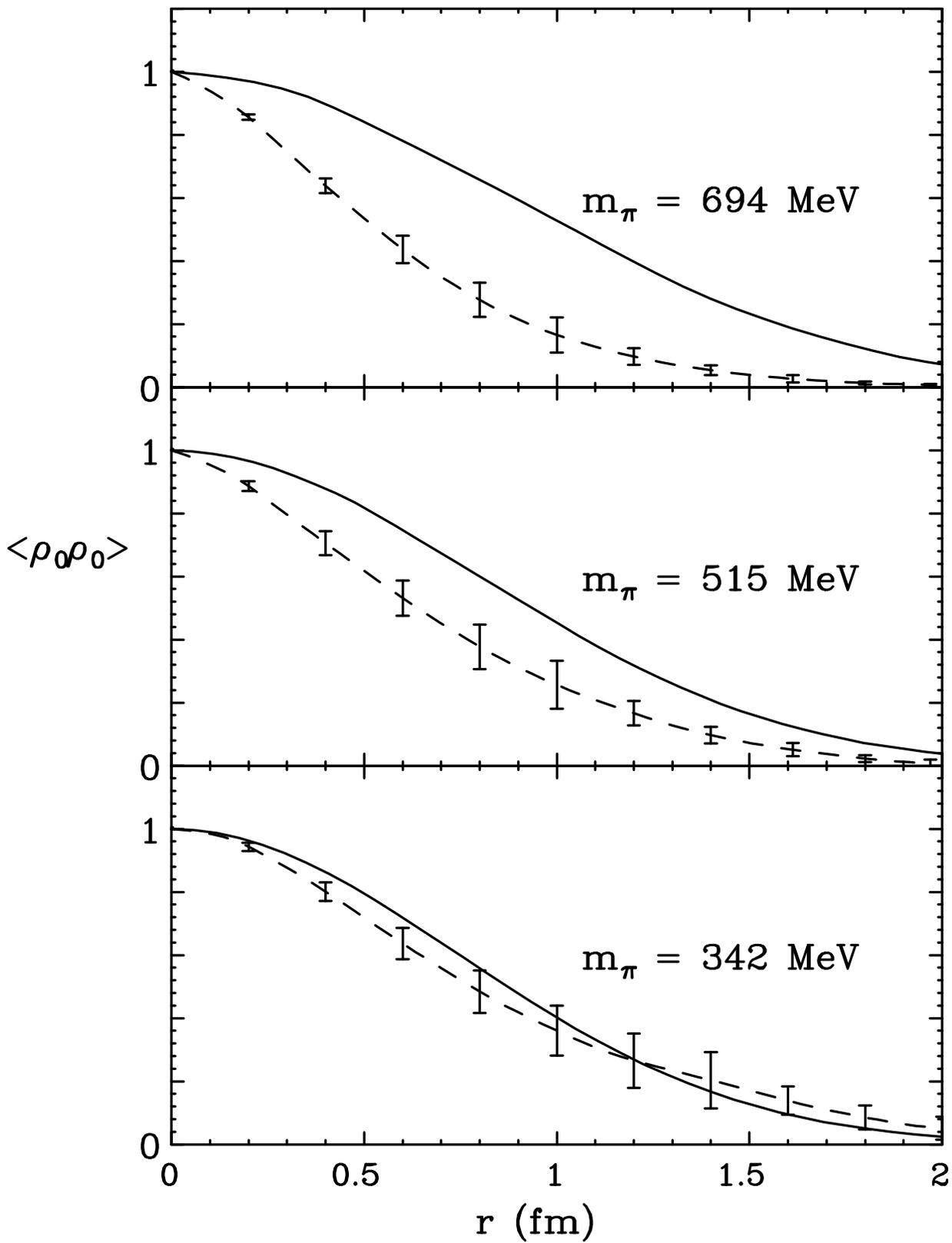
\item{Fig. 4} Comparison of density-density correlation functions
on the lattice and in the Skyrme model. The solid curves denote the
Skyrme results for the parameters in Table I corresponding to each
lattice quark mass.  Lattice measurements of the correlation functions
are shown by error bars joined by dashed lines to guide the eye,
and all correlation functions are normalized to one at the origin.
\medskip
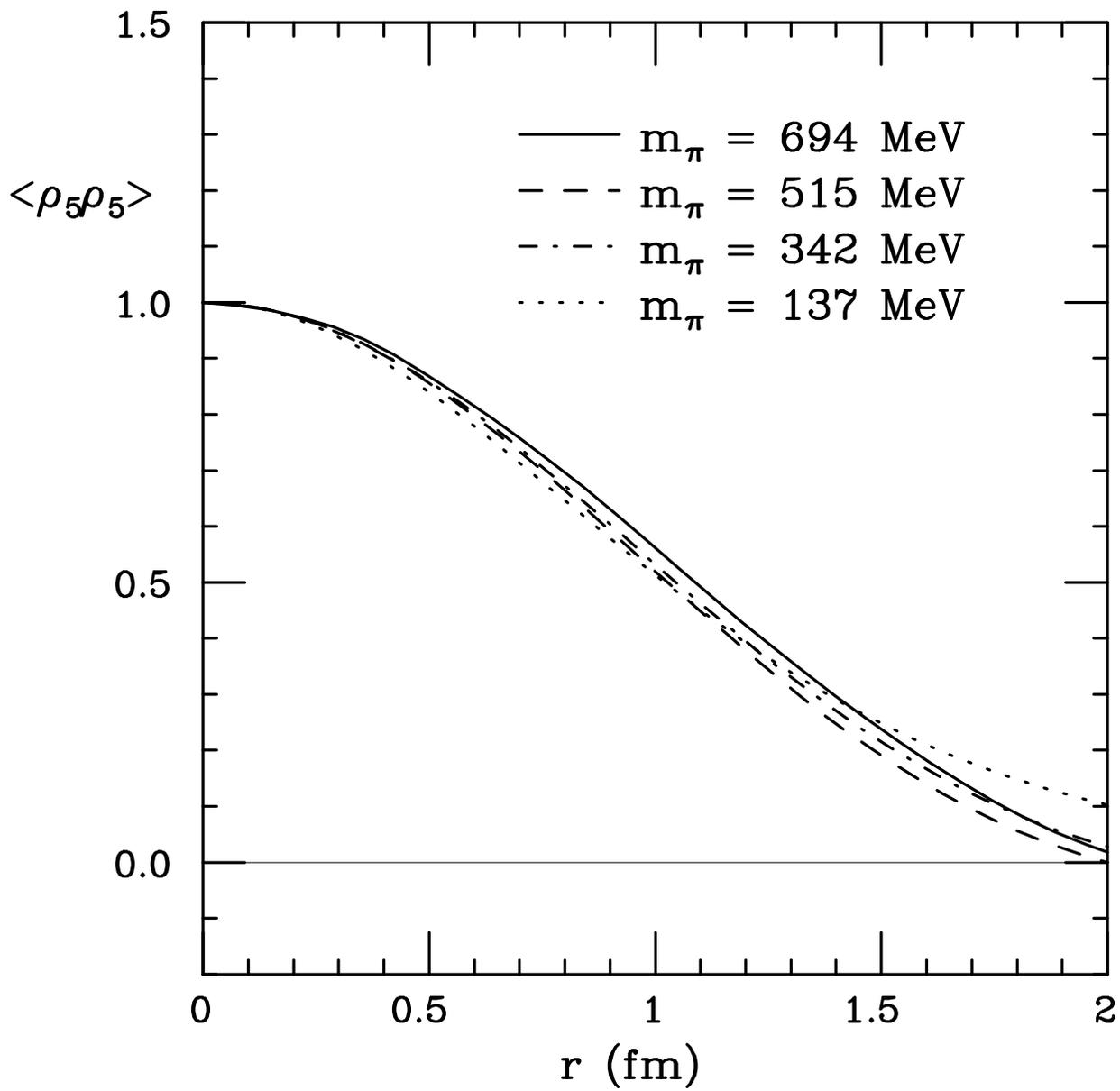
\item{Fig. 5} Mass dependence of the pseudoscalar correlation function
in the Skyrme model.  The four curves show the spatial distribution of
the correlation function for each of the four sets of parameters in
Table I.
\medskip
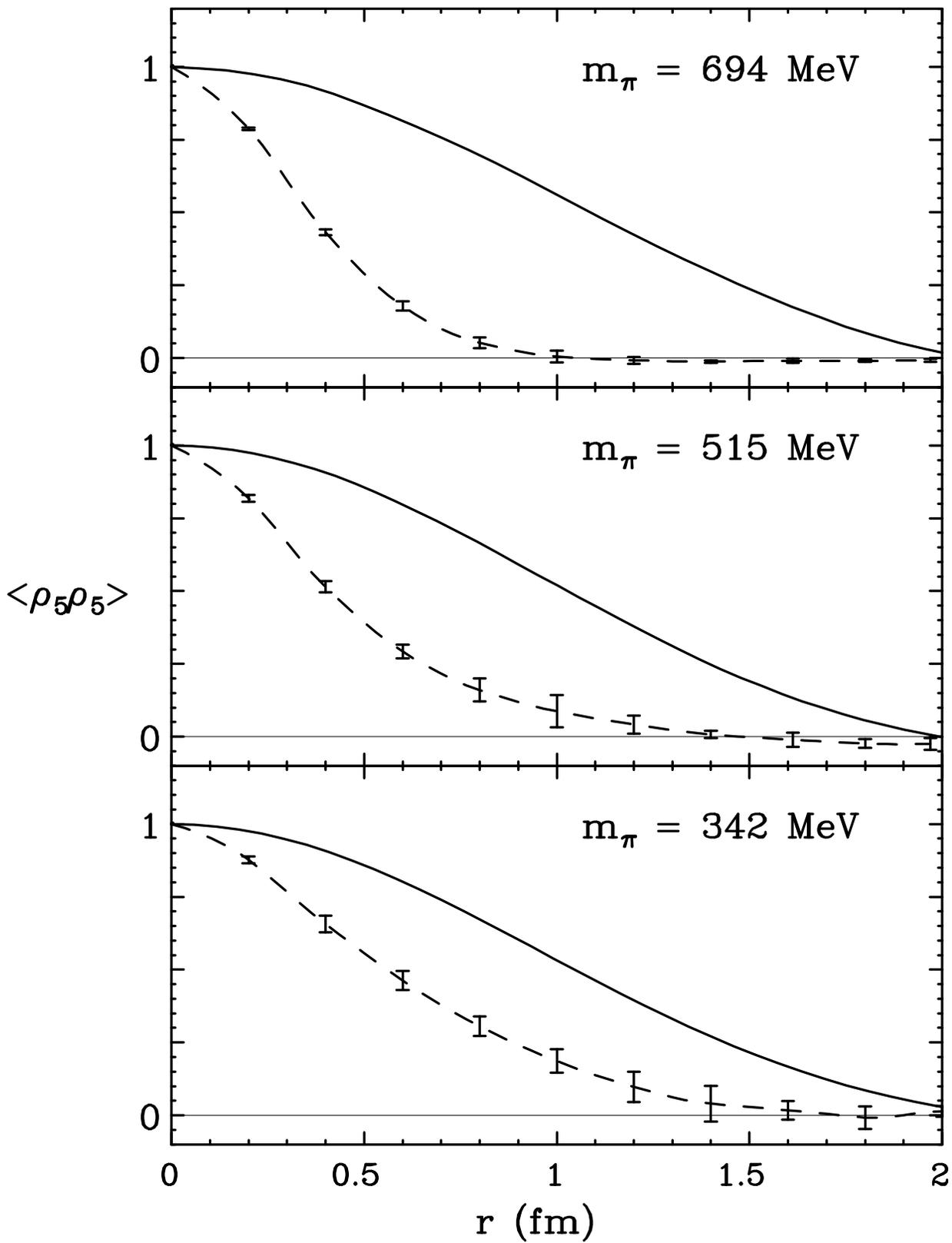
\item{Fig. 6} Comparison of pseudoscalar correlation functions
on the lattice and in the Skyrme model. The solid curves denote the
Skyrme results for the parameters in Table I corresponding to each
lattice quark mass.  Lattice measurements of the correlation functions
are shown by error bars joined by dashed lines to guide the eye,
and all correlation functions are normalized to one at the origin.

\vfill\eject

\centerline{\hbox{
\vbox{\offinterlineskip
\hrule height .8pt
\halign{
{\vrule height 12pt depth 6pt width 0pt}\vrule width .8pt#&
\hfil\enskip#\enskip\hfil&\vrule#&
\hfil\enskip#\enskip\hfil&\vrule#&
\hfil\enskip#\enskip\hfil&\vrule#&
\hfil\enskip#\enskip\hfil&\vrule#&
\hfil\enskip#\enskip\hfil&\vrule width .8pt#&
\hfil\enskip#\enskip\hfil&\vrule#&
\hfil\enskip#\enskip\hfil&\vrule#&
\hfil\enskip#\enskip\hfil&\vrule width .8pt#\cr
& \multispan{15}{\hfil {\bf TABLE I} \hfil} &\cr
\noalign{\hrule height .8pt}
& \multispan{9}{\hfil Skyrme \hfil} && \multispan{5}{\hfil Lattice \hfil} &\cr
\noalign{\hrule height .8pt}
& $\kappa$ && && && && && {\bf 0.167} && {\bf 0.1639} && {\bf 0.16} &\cr
& $m_q$ (MeV) && && 40 && 95 && 175 && 40 && 95 && 175 &\cr
& $m_\pi$ (MeV) && {\bf 137} && {\bf 342} && {\bf 515} && {\bf 694}
              && 340(7) && 511(5) && 691(3) &\cr
& $e$ && 4.83  && 3.79 && 3.23 && 2.73 && && && &\cr
& $F_\pi$ (MeV) && 108.1 && 60.32 && 46.87 && 34.87 && && && &\cr
& $(eF_\pi)^{-1}$ (fm) && 0.383 && 0.862 && 1.321 && 2.141 && && && &\cr
& $\beta = m_\pi / (e F_\pi)$ && 0.262 && 1.497 && 3.402 && 7.290 && && && &\cr
& $m_N$ (MeV) && {\bf 939} && {\bf 915} && {\bf 1097} && {\bf 1321}
                        && 915(6) && 1097(11) && 1321(10) &\cr
& $m_\Delta$ (MeV) && {\bf 1232} && {\bf 1091} && {\bf 1223} && {\bf 1403}
                 && 1091(23)$^*$ && 1223(11) && 1403(10) &\cr
& $r_S$ (fm)  && 0.690 && 0.917 && 0.970 && 1.075 && && && &\cr
& $r_V$ (fm)  && 1.048 && 1.088 && 1.132 && 1.256 && && && &\cr
& ${9 \over 8}r_S - {1 \over 8}r_V$ (fm) && 0.632 && 0.894 && 0.947 && 1.050
            && 1.0(5) && 0.79(20) && 0.75(15) &{\vrule depth 8pt width 0pt}\cr
}\hrule height .8pt}}}

\bigskip
\midinsert
\narrower
\noindent
{\bf Table I.~~} Parameters of the Skyrme model for lattice calculations
corresponding to three different bare quark masses and for physical
hadrons. The input hadron masses are shown in bold face font and all other
quantities are derived from them. The isoscalar and isovector r.m.s. radii are
denoted $r_S$ and $r_V$ respectively. The first column uses physical hadron
masses and the next three columns use the lattice hadron masses given in the
lattice data shown at the right.  The lattice data are taken from
Refs.~[7-10], with the exception of $m_\Delta$ at $\kappa=0.167$ denoted by an
$*$ which was taken from Ref.~12. For reference, we have also tabulated the
values of the bare quark mass
$m_q \equiv {1\over{2\kappa}} -{1\over{2\kappa_c}}$
corresponding to the values of the hopping parameter $\kappa$ for each lattice.
\endinsert

\bye